\documentclass{article}

\usepackage{epsf,a4wide,cite}

\usepackage[tbtags]{amsmath}
\usepackage{feynmp,subfigure}

\begin{document}

\hbox{}
\nopagebreak
\vspace{-3cm}
\begin{flushright}
{\sc OUTP-99 22P} \\
\end{flushright}

\vspace{1in}

\begin{center}
{\Large \bf Vector potential versus colour charge density in low-x evolution}
\vspace{0.1in}

\vspace{0.5in}
{\large  Alex Kovner and J. Guilherme Milhano}\\
{\small
Theoretical Physics, University of Oxford, 1 Keble Road, Oxford,
OX1 3NP, UK}\\

\vspace{0.5in}

{\bfseries\sc  Abstract}\\
\end{center}
\vspace{.2in}
We reconsider the evolution equations for multigluon correlators
derived in \cite{moap}.
We show how to derive these equations directly in terms of vector
potentials (or colour field strength) avoiding the introduction of the
concept of colour charge density in the intermediate steps.
The two step procedure of deriving the evolution of the charge density
correlators followed by the solution of classical equations for the
vector potentials is shown to be consistent with direct derivation of
evolution for vector potentials.
In the process we correct some computational errors of \cite{moap} and
present the corrected evolution equations which have a somewhat
simpler appearance.

\vfill

\newpage

\section{Introduction}

With the advent of HERA there has been a great increase in the scope
of the theoretical effort to understand the physics of hadronic
scattering at high energy.
This is a challenging subject especially as it might provide a
bridge between perturbative partonic physics of short distance
processes (e.g. DIS at moderate $x_{Bj}$) and soft physics of hadronic
states which, presumably, dominate the high energy asymptotics.
The borderline between the two --- the "semihard" physics --- is
interesting also in its own right.
In essence it is the physics of partonic systems which are, on one hand
dense enough for new collective phenomena to play important role but,
on the other hand are perturbative since the average momentum transfer
between the partons is high enough.
In this semihard regime one expects to see perturbativelly
controllable nonlinear effects which depart from the standard linear
evolution of DGLAP \cite{dglap} or BFKL \cite{bfkl} type and
subsequently lead to unitarization of hadronic cross sections.

An approach to this high partonic density regime from the partonic
side has been spearheaded by Levin and collaborators \cite{levin}
based on earlier work by Mueller \cite{mueller} and is an "all
twist" generalization of the GLR recombination picture \cite{glr,mq}. 
It lead to the formulation of a nonlinear evolution equation which
exhibits a perturbative mechanism of unitarization. 
Analysis of this equation suggests that already at present HERA
energies the nonlinearities in the gluon sector are considerable and
linear evolution for gluons should break down. 
Better experimental data on gluon distributions would be extremely
valuable in order to verify/falsify this assessment.

A complementary approach was pioneered some years ago by McLerran and
Venugopalan \cite{mv}.
It was later somewhat reshaped conceptually and considerably developed
technically in a series of papers \cite{jkmw,jklw,moap,soap}.
Here the idea is that in the high density regime rather than using
partonic language, it is more appropriate to use the language of
classical fields.
The hadron then is considered as an ensemble of configurations of the
gluon field.
The statistical weight that governs the contributions of different
configurations to the ensemble averaging changes when one probes the
hadron on different time scales.
Decreasing $x_{Bj}$ corresponds to increasing the time resolution and
therefore corresponds to probing the hadron on shorter time scales.
The change of the statistical weight with $x_{Bj}$ is governed by the
evolution equation.
As long as the field intensity is large enough this evolution should
be perturbative in $\alpha_s$ but essentially nonlinear in the field
intensity itself.

This evolution equation to first order in $\alpha_s$ was derived in
\cite{jklw,moap}.
We will refer to it as the JKLW equation in the following. 
In \cite{soap} the double logarithmic limit of this evolution was
considered. 
It was shown that in this limit the evolution of the gluon density
becomes unitary at large density. 
Qualitatively, the evolution is very similar to that discussed in
\cite{levin}, although the details are different. 
A detailed numerical study of the doubly logarithmic limit of the JKLW
evolution was recently performed in \cite{jw}.

Technically the derivation of \cite{moap} is fairly involved. 
Several consistency checks were performed in \cite{jklw} and
\cite{soap} to make sure that the known results are recovered from the
general evolution equation in the weak field limit.
This includes the BFKL equation, the doubly logarithmic approximation
to the DGLAP equation and the GLR equation.
It is, however, desirable to have some additional independent checks on
the equation which do not involve the weak field limit. 
It is the aim of the present paper to provide one such check.

In a nutshell the issue we address is the following. 
The evolution equation in \cite{jklw,moap,soap} was derived invoking a
two step procedure.
Rather than considering directly the evolution of the correlators of
the gluon field, one first considers the evolution of the colour
charge density. 
In the second step, one re-expresses the evolution equation for the
charge density as the evolution equation for the vector potential
(chromoelectric field).
The evolution of the field correlators is in fact what one is after,
since it is the vector potential and not the charge density that
couples directly to fermions and, therefore, are more directly related
to physical observables.
Our observation in this paper is that one can avoid the introduction
of the colour charge density altogether and derive the evolution
equations directly for the field correlators.
This procedure has the advantage of being somewhat simpler both
technically and conceptually.
Nevertheless, the final result for the evolution should be the same as
in the two step procedure of \cite{jkmw,jklw,moap,soap}. 
The comparison of our results with the earlier derived formulae
provides us with a consistency check on the calculation.

We find, in fact, that the results presented in \cite{moap} are not
entirely correct.
However, after correcting some algebraic mistakes in \cite{moap} we
show explicitly that the two approaches yield identical results. 
We provide the corrected expressions for the "kernels" of the
evolution equation, which are somewhat simpler than the expressions
found in \cite{moap}.
We also clarify the issue of possible Gribov ambiguity and show
explicitly that the divergent Jacobians, which appeared in the
intermediate steps of the derivations in \cite{soap}, cancel completely
in the final expressions for the correlators of the chromoelectric
field.
Therefore, the Gribov ambiguity, although affecting the relation
between the colour charge density and the chromoectric field, does not
affect the evolution of the field correlators, at least to order
$\alpha_s$. 
Since the procedure discussed in the present paper avoids the
introduction of the charge density entirely, the whole approach is
free from the Gribov problem.

Perhaps somewhat surprisingly the corrections to the results of
\cite{moap} that we find do not affect either the weak field limit
discussed in \cite{jklw} or the doubly logarithmic limit of
\cite{soap}. 
They therefore have no bearing on the derivation of the BFKL equation
in our approach and also do not help to reconcile the doubly
logarithmic limit of the JKLW equation \cite{soap} with the nonlinear
equation studied in \cite{levin}.

The plan of the paper is the following. In Section 2 we briefly recap
the procedure of the derivation of the evolution equation as described
in \cite{jklw,moap,soap} and reformulate it directly in terms of the
gluon field correlators.
In Section 3, using some of the results of \cite{moap}, we calculate
the real and virtual parts of the evolution in terms of the field
correlators and provide the corrections to the results of \cite{moap}.
Finally, in Section 4 we discuss our results.

\section{The JKLW evolution}

First, let us briefly recall the framework and the results of
\cite{jklw, moap,soap}.
In this approach the averages of gluonic observables in a hadron are
calculated via the following path integral
\begin{multline}
        <O(A)>  = \int D\rho DA^\mu O(A)
                \exp\Big\{ -\int d^2 x_\perp F[\rho ^a(x_\perp)]
                -i\int d^4 x \frac{1}{4}{\rm tr} F^{\mu\nu}F_{\mu\nu} \\
        +{\frac{1}{N_c}} \int d^2 x_\perp dx^- \delta (x^-)
                \rho_{a}(x_\perp) {\rm tr}T_a W_{-\infty,\infty} 
                [A^-](x^-,x_\perp)\Big\}
\label{action}
\end{multline}
where the gluon field strenght tensor is given by
\begin{equation}
        F^{\mu\nu}_{a} = 
        {\partial}^{\mu} A^{\nu}_{a} - {\partial}^{\nu} A^{\mu}_{a} - 
        g f_{abc}A^{\mu}_{b}A^{\nu}_{c}
\end{equation} 
and $W$ is the Wilson line in the adjoint representation along the
$x^+$ axis
\begin{equation}
        W_{-\infty ,+\infty}[A^{-}](x^- , x_{\perp}) = 
        P\exp\Bigg[+i g \int dx^+ A^-_a (x^+,x^- , x_{\perp}) T_a \Bigg] 
\end{equation}
  
The hadron is represented by an ensemble of colour charges localized
in the plane $x^-=0$ with the (integrated across $x^-$) colour charge
density $\rho(x_\perp)$.  The statistical weight of a configuration
$\rho(x_\perp)$ is
\begin{equation}
        Z=\exp \{-F[\rho]\}
\label{z}
\end{equation}
In the tree level approximation (in the light cone gauge $A^+=0$) the
chromoelectric field is determined by the colour charge density
through the equations
\begin{equation}
        F^{+i}={1\over g}\delta(x^-)\alpha^{i}(x_\perp)
\label{chrom}
\end{equation}
and the two dimensional vector potential $\alpha^i(x_\perp)$ is "pure
gauge", related to the colour charge density by
\begin{align}
        {\partial}^{i}{\alpha}_{a}^{j} &- 
        {\partial}^{j}{\alpha}_{a}^{i} - 
        f_{abc}{\alpha}_{b}^{i}{\alpha}_{c}^{j}=0 \nonumber \\
        {\partial}^{i}{\alpha}_{a}^{i}  &
        =-{\rho}_{a}
\label{sol}
\end{align}

Integrating out the high longitudinal momentum modes of the vector
potential generates the renormalization group equation, which has the
form of the evolution equation for the statistical weight $Z$
\cite{jklw} 
\footnote{All the functions in the rest of this paper
depend only on the transverse coordinates. For simplicity of notation
we drop the subscript $\perp$ in the following.}
\begin{equation}
        {d\over d\zeta}Z= \alpha_s \left\{{1\over 2}{\delta^2
        \over\delta\rho(u)\delta\rho(v)}\left[Z\chi(u, v) \right] -
        {\delta\over\delta\rho(u)}\left[Z\sigma(u)\right]\right\}
\label{final}
\end{equation}
In the compact notation used in Eq.~(\ref{final}), both $u$ and $v$
stand for pairs of colour index and transverse coordinates, with
summation and integration over repeated occurrences implied. The
evolution in this equation is with respect to the rapidity $\zeta$,
related to the Feynman $x$ by
\begin{equation}
        \zeta=\ln 1/x
\end{equation}
Technically it arises as a variation of $Z$ with the cutoff imposed on
the longitudinal momentum of the fields $A^\mu$.  
The quantities $\chi[\rho]$ and $\sigma[\rho]$ have the meaning of the
mean fluctuation and the average value of the extra charge density
induced by the high longitudinal momentum modes of $A^\mu$. 
They are functionals of the external charge density $\rho$.
The explicit expressions have been given in \cite{moap} and it is our
aim in this paper to provide a check on these expressions.

Eq.~(\ref{final}) can be written directly as an evolution equation for
the correlators of the charge density. 
Multiplying Eq.~(\ref{final}) by $\rho(x_1)...\rho(x_n)$ and
integrating over $\rho$ yields
\begin{multline}
        {d\over d\zeta}<\rho(x_1)...\rho(x_n)>= \\ 
        = \alpha_s \Bigg[\sum_{0<m<k<n+1}
        <\rho(x_1)...\rho(x_{m-1})\rho(x_{m+1})... 
        \rho(x_{k-1})\rho(x_{k+1})...\rho(x_n)\chi(x_m, x_k)> \\ 
        +\sum_{0<l<n+1}<\rho(x_1)...\rho(x_{l-1})\rho(x_{l+1})
        ...\rho(x_n)\sigma(x_l)>\Bigg]
\label{correl}
\end{multline}

This set of equations for the correlators of the colour charge density
completely specifies the evolution of the hadronic ensemble as one
moves to higher energies (or lower values of $x$).

The evolution equations for the correlators of the charge density can
be rewritten as equations for the correlators of the vector potential
\cite{soap}.
\begin{multline}
\label{correlf}
        {d\over d\zeta}
        <\alpha_{a_1}^{i_1}(x_1)...\alpha_{a_n}^{i_n}(x_n)> = \\  
        = \alpha_s \Bigg[\sum_{0<l<n+1}
        <\alpha_{a_1}^{i_1}(x_1)...
        \alpha_{a_{l-1}}^{i_{l-1}}(x_{l-1})
        \alpha_{a_{l+1}}^{i_{l+1}}(x_{l+1})...
        \alpha_{a_n}^{i_n}(x_n)\sigma_{a_l}^{i_l}(x_l)> \\  
        + \sum_{0<m<k<n+1}<\alpha_{a_1}^{i_1}(x_1)...
        \alpha_{a_{m-1}}^{i_{m-1}}
        (x_{m-1})\alpha_{a_{m+1}}^{i_{m+1}}(x_{m+1})... \\ 
        \times
        \alpha_{a_{k-1}}^{i_{k-1}}(x_{k-1})
        \alpha_{a_{k+1}}^{i_{k+1}}(x_{k+1})...
        \alpha_{a_n}^{i_n}(x_n)\chi_{a_ma_k}^{i_mi_k}(x_m, x_k)>
        \Bigg]
\end{multline}
The quantities $\chi_{ab}^{ij}$ and $\sigma_a^i$ have a very simple
physical meaning.
The high momentum modes of the vector field which have been integrated
out in order to arrive at the evolution equation induce extra colour
charge density $\delta\rho$.  
The average value of this induced density and its mean fluctuation
appear in the evolution equations eq.(\ref{correl}) as $\sigma_a^{ }$
and $\chi_{ab}^{  }$.  
The appearance of the induced colour charge density leads to the
change in the value of the chromoelectric field through the solution
of Eq.~(\ref{sol}) with $\rho+\delta\rho$ on the right hand side. 
The quantities $\sigma_a^i$ and $\chi_{ab}^{ij}$ 
are the average value
and the mean fluctuation of the induced field respectively.

It is perhaps helpful to explain how  $\sigma_a^i$ and
$\chi_{ab}^{ij}$ 
were obtained
in
\cite{soap}.
As shown in \cite{moap}, the induced charge density can be decomposed
into two pieces
\footnote{The reason for the notation $\tilde\rho$  rather than simply
$\rho$ will be explained in the next section.}
\begin{equation}
        \delta\rho=\delta\tilde\rho_1+\delta\tilde\rho_2
\end{equation}
The first piece $\delta\tilde\rho_1$ is order $g$ while the second
piece $\delta\tilde\rho_2$ is order $g^2$. 
The $\delta\tilde\rho_1$ is time dependent, and
has zero average value, while its mean fluctuation is order $g^2$.
The $\delta\tilde\rho_2$ being $O(g^2)$ contributes only to the
average value of $\delta\tilde\rho$ and not to the mean fluctuation.
Assuming that the classical equations eq.(\ref{sol}) hold not only for
the background field but also for the relevant part of the fluctuation
field one can solve those equations perturbatively.
Writing
\begin{equation}
        \delta\alpha^i=\delta\alpha_1^i+\delta\alpha_2^i
\end{equation}
with $\delta\alpha_1$ being $O(g)$ and $\delta\alpha_2$ being $O(g^2)$
and keeping in the classical equations all terms to order $g^2$ we
have
\begin{align}
        &D_{ab}^i\delta\alpha_{1b}^j
        -D_{ab}^j\delta\alpha_{1b}^i
        +D_{ab}^i\delta\alpha_{2b}^j
        -D_{ab}^j\delta\alpha_{2b}^i
        -f_{abc}\delta\alpha_{1b}^i\delta\alpha_{1c}^j=0\nonumber \\
        &\partial^i\delta\alpha_{1a}^i+\partial^i\delta\alpha_{2a}^i
        =-(\delta\tilde\rho_{1a}+\delta\tilde\rho_{2a})
\label{solp}
\end{align}
We have defined for convenience
\begin{align}
        \alpha_{ab}^i &= f_{abc}\alpha^i_c \nonumber \\
        D_{ab}^i &= \partial^i\delta_{ab}+\alpha^i_{ab}
\label{def}
\end{align}

To order $g$ we find
\begin{equation}
        \delta\alpha^i_1=-D^i{1\over \partial D}\delta \tilde\rho_1
\label{corresp1}
\end{equation}
Therefore, to order $g^2$
\begin{equation}
        \chi_{ab}^{ij}(x,y)=
        r_{ac}^i(x,u)\chi_{cd}^{  }(u,v)r_{db}^{\dagger j}(v,y)
\label{indcor}
\end{equation}
with
\begin{equation}
        r_{ab}^i(x,y)=
        - <x|[D^i\frac{1}{\partial D}]_{ab}|y>
\label{r}
\end{equation}
Here $<x|O|y>$ denotes a configuration space matrix element in the
usual sense.

At order $g^2$ we have
\begin{equation}
        \delta\alpha^i_2=-D^i{1\over \partial D}
        \delta \tilde\rho_2-{1\over 2}\epsilon^{ij}\partial^j
        {1\over D\partial}\delta\alpha_{1}\times\delta\alpha_{1}
\label{corresp2}
\end{equation}
Here the cross product is defined as 
\begin{equation*}
        A\times B = f_{abc}\epsilon^{ij}A_a^iB_b^j
\end{equation*}

We thus have
\begin{equation}
        \sigma_{a}^{i}(x)=
        r_{ab}^i(x,u)\sigma_{b}^{ }(u)+p_{abc}^i(x,u,v)\chi_{bc}^{  }(u,v)
\label{indfield}
\end{equation}
with
\begin{equation}
        p^i_{abc}(x,y,z)=
        - \frac{1}{2}(\epsilon^{ij}\partial^j[\frac{1}{D\partial}]_{ad})(x,u)
        f_{dfe}\epsilon^{kl} r^k_{fb}(u,y)r^l_{ec}(u,z)
\label{p}
\end{equation}

The procedure of deriving eq.(\ref{correlf}) employed in
\cite{moap,soap} consists, therefore, of two steps. 
One first splits the gluon field into the classical background field
$\alpha^\mu$ and the fluctuation field $a^\mu$. 
The modes of the fluctuation field with longitudinal momenta in some
range $\alpha_s\ln {\Lambda^+\over \Lambda^{'+}}$ are assumed to be
small. 
One defines operatorialy the induced charge density $\delta\rho$ in
terms of the fluctuation fields $a^\mu$ and the quantities $\sigma$
and $\chi$ are calculated by integrating out the fluctuation fields
perturbatively.
In the second step, one solves classical equations of motion which
include
the induced charge density and calculates $\sigma^i$ and $\chi^{ij}$. 

Clearly, consistency requires that the two step procedure that leads
from eq.(\ref{action}) through eqs.(\ref{final},\ref{correl}) to the
evolution equations eq.(\ref{correlf}, \ref{indcor}, \ref{indfield})
must be equivalent to the following.
Start with the equivalent of eq.(\ref{action})
\footnote{We note that the statistical weight $Z[\alpha_i]$ which
appears in eq.(\ref{actionf}) is not equal to $Z$ of eq.(\ref{z})
since going from eq.(\ref{action}) to eq.(\ref{actionf}) involves the
change of variables $\rho\rightarrow\alpha_i$. The two statistical
weights, therefore, differ by an appropriate Jacobian.}
\begin{multline}
        <O(A)> =\int D\alpha^i DA^\mu O(A) Z[\alpha_i(x_\perp)]
        \exp\{ -i\int d^4 x {1\over 4}{\rm tr} F^{\mu\nu}F_{\mu\nu} \\
        -{{1}\over{N_c}} \int d^2 x_\perp dx^- \delta (x^-)
        \partial^i\alpha_a^i(x_\perp) {\rm tr}T_a W_{-\infty,\infty} 
        [A^-](x^-,x_\perp)\} 
\label{actionf}
\end{multline}
Integrate out the high longitudinal momentum components of $a^\mu$ as
before, but instead of calculating the induced charge density
$\sigma_a^{ }$ and $\chi_{ab}^{  }$, calculate directly the induced
chromoelectric field $\sigma_a^i$ and $\chi_{ab}^{ij}$. 
Technically this calculation is somewhat simpler, since there is no
need to consider the operator $\delta\rho$, which is nonlinear in the
fluctuation field $a^\mu$. 
Instead, one directly calculates the distribution of the static
component of $a^\mu$.
The resulting evolution equations should coincide with
eq.(\ref{correlf}).

With this formulation one circumvents completely the need to introduce
the colour charge density $\rho$ and to solve classical equations 
for $\alpha^i$ in terms of $\rho$.
While one may want to introduce $\rho$ for reasons of convenience, our
present understanding is that it is not necessary from the point of
view of physics. 
The physics that our approach is meant to address
is that of the evolution of the hadronic ensemble.
The relation between $\alpha^i$ and $\rho$ on the other hand is
supposed to hold at every value of $\zeta$, and therefore itself is
unrelated to evolution in $\zeta$. 
The concept of $\rho$ may be sometimes useful to formulate models for
the statistical weight $Z$ at some particular value of $\zeta$ as was
the original motivation of \cite{mv}. 
This could then serve as an initial condition for the evolution. 
This, however, is a separate question and we do not intend to address it
here.

Before we proceed further, we wish to make one more comment about the
relation between the chromoelectric field and the colour charge
density eqs.(\ref{corresp1},{\ref{corresp2}).
Both these equations contain the dangerous factor $(\partial
D)^{-1}$. 
The operator $\partial D$ has zero, as well as negative eigenvalues
and is very reminiscent of the operators usually associated with
the Gribov ambiguity in nonabelian gauge theories. 
In fact, it is quite clear that it has precisely the same origin. 
The second equation in eq.(\ref{solp}) has the form of the Lorentz
like gauge fixing condition on the fluctuation field $\delta\alpha$. 
Since the calculation is performed in a nonvanishing background field,
the Lorentz gauge indeed suffers from Gribov ambiguity precisely due
to negative eigenvalues of the operator $\partial D$.

Given this, one may worry that our perturbative calculation is plagued
with the Gribov ambiguity
\footnote{In standard perturbation theory, the Gribov ambiguity does
not show up in any finite order. This is due to the fact that one
expands the operator $\partial D$ and its inverse in powers of the
coupling constant. To leading order then the operator does not have
any negative eigenvalues, which ensures that no problems arise in
finite order perturbative calculations. Our situation is, however,
different. Since our background field is not assumed to be $O(g)$, the
operator cannot be expanded. Therefore, there is no guarantee that the
problem does not show up even in perturbation theory.}.
However this is not necessarily the case. 
The point is that $\delta\rho$ itself is not arbitrary. 
It is calculated through the fluctuation field and, at the end of the
day, is averaged over with some statistical weight
$Z[\delta\rho]$. 
It could well be that the statistical weight is such that it only
allows induced charge density of the form $\delta\rho=\partial D X$
with regular $X$. 
If that is the case, the dangerous denominator cancels and the induced
field is well defined and regular. 
In fact, in our present formulation where the calculation is performed
directly in terms of the field, it is almost clear that this should
indeed happen. 
In this setup one calculates directly $\delta\alpha$, and
eqs.(\ref{corresp1},\ref{corresp2}) should be read from right to left,
as equations determining an auxilliary quantity $\delta\rho$ through
$\delta\alpha$ rather than the other way round. 
The operator $\partial D$ then appears in the numerator and all
expressions are regular.
In fact we will show in the next section by explicit calculation that
all ``dangerous factors'' indeed cancel in the final expressions for
$\chi_{ab}^{ij}$ and $\sigma_a^i$.

Note that, if one insists on formulating the problem in terms of the
colour charge density, the absence of the Gribov ambiguity implies a
nontrivial consistency condition on the statistical weight $Z[\rho]$. 
Taking an arbitrary weight $Z$ will render the calculation of
chromoelectric field correlators ill defined especially at strong
fields (strong coupling). This was indeed observed in the numerical
calculation \cite{gv} where a simple Gaussian in $\rho$ was used as the
weight function
\footnote{This problem does not arise in the more recent numerical work
\cite{kv} since in effect this work uses a different definition of
$\rho$ for which the relations analogous to
eqs.(\ref{corresp1},\ref{corresp2}) do not involve singular factors.}.

In the next section we will calculate $\sigma_a^i$ and 
$\chi_{ab}^{ij}$ induced by high longitudinal momentum modes.

\section{The Induced Chromoelectric Field}

The main ingredients needed for the calculation of the induced
chromoelectric field are the eigenfunctions of the quadratic action
for the small fluctuations in the static background $\alpha^i$.
Solving the classical equations of motion that follow from the action eq.(\ref{actionf}) at fixed $\alpha^i$ we find 
(in the gauge $\partial^i A^i(x^+\rightarrow -\infty)=0$) the classical solution

\begin{equation}
  A_{cl}^{-}=0, \quad  A^i_{cl} =\alpha^i (x_\perp)\theta(x^-)
\end{equation}

Defining the quantum fluctuation field $a^\mu$ by
 $A^\mu =A^\mu_{cl}+a^\mu$ and expanding the action to 
second order in $a^\mu$ we have
\begin{equation} 
        S=\frac{1}{2g^2} \Bigg\{ 
                a^-_{x} K_{xy} a^-_{y} +
                2a^-(\partial^+ Da - 2 fa ) +
                2\partial^+a^i\partial^-a^i + 
                a^i \bigg[ D^2 \delta^{ij} - D^{i}D^{j} \bigg] a^j
        \Bigg\}
\label{eq:action}
\end{equation} 
Here we are using the notation 
\begin{align}
        &[fa]_{a}(x^+,x^-,x_\perp ) = 
                \delta (x^-) \alpha^i_{ab}(x_\perp )
                a^i_{b}(x^+,x^-,x_\perp ) \nonumber \\ 
        &Da =
                D^{i}[\alpha]a^{i}=(\partial^i\delta_{ab}+
                \theta(x^-)\alpha^i_{ab})a_b^i 
\label{cov}
\end{align} 
and as previously
\begin{equation}
        \alpha^i_{ab}=f_{abc}\alpha^i_c
\end{equation}

The operator $K$ is
\begin{equation} 
        K_{ab}^{xy} = 
                - \bigg[(\partial^{+})^2 \delta_{ab} 
                +\partial^i\alpha^i_{ab} \delta (x^-) \frac{1}{\partial^-} \bigg] 
\label{K}
\end{equation} 
Note that there is no ambiguity in the definition of the operator
$1/\partial^- $ in this expression. It is defined in the sense of
principal value.
This follows directly from the fact that the matrix $\alpha^i_{ab}$ is
antisymmetric and therefore the term involving $1/\partial ^-$ in
eq.(\ref{eq:action}) vanishes for zero frequency fields.

This eigenfunctions of the quadratic fluctuation operator
have been found
in \cite{moap} and we cite here
the relevant results.
The calculation is performed in the lightcone gauge $A^+=0$ with the
residual gauge freedom fixed by the condition
\begin{equation}
        \partial^i A^i(x^-\rightarrow -\infty)=0
\end{equation}

It is convenient to define an auxiliary field
\begin{equation}
        \tilde a^- = a^-+ K^{-1} (\partial^+ Da - 2 fa )
\label{tildea}
\end{equation}
This field can be seen to decouple from $a_i$. Its correlator is 
\begin{equation}
        <\tilde a^-_x\tilde a^-_y>=K^{-1}_{x,y}
\end{equation}
The operator $K$ Eq.(\ref{K}) has zero modes. 
Defining the projector matrices
$\eta$ and $\mu$ by 
\begin{equation}
        \mu_{ab} \partial^i\alpha^i_{bc} \frac{1}{\partial^-}  = 0,\quad 
        \eta_{ab} \partial^i\alpha^i_{bc} \frac{1}{\partial^-} = 
                \rho_{ac} \frac{1}{\partial^-} 
\end{equation}
and
\begin{equation}
        \mu + \eta = 1 ,\quad
        \mu^{2} = \mu ,\quad 
        \eta^{2} = \eta
\label{eq:project}
\end{equation}
we can write the normalizable zero modes of $K$ in the form
\begin{equation}
        f_a(x_\perp ,x^-,p^-)=\mu_{ab}f(x_\perp ,p^-)
\label{zero}
\end{equation}
The operator $K$ is therefore, strictly speaking, non invertible. 
The operator $K^{-1}$ in eq.(\ref{tildea}) has to be understood as the
inverse of $K$ on the space of functions which does not include the
functions Eq.(\ref{zero}). 
Further, it is only the nonzero mode part of $a^-$ that enters the
definition of $\tilde a^-$ in eq.(\ref{tildea}).

For our calculation we will need the properly normalized solutions of
the equations of motion that follow from the action
eq.(\ref{eq:action}).
The complete set of these solutions was found in \cite{moap}
\begin{multline}
        a^{i}_{p^-,r} =  
        ge^{ip^{-}x^{+}} \int d^{2}p_{\perp} \bigg[ \theta (-x^-)
        \exp\left( i{p^{2}_{\perp} \over 2p^{-}}x^{-} - 
        ip_{\perp}x_{\perp}\right) v^{i}_{-, r}(p_{\perp}) \\
        + \theta (x^{-}) U (x_\perp ) 
        \exp\left( i{p^{2}_{\perp} \over 2p^{-}}x^{-} - 
        ip_{\perp}x_{\perp}\right)
        \left[U^{\dagger}v^{i}_{+,r}\right](p_\perp ) 
        + \theta (x^{-}) \gamma^{i}_{+,r}\bigg] 
\label{solut}
\end{multline}
The frequency $p^-$ is a good quantum number since the background
field is static.
Here $r$ is the degeneracy label, which labels independent solutions
with the frequency $p^-$. 
In the free case it is conventionally chosen as the transverse
momentum, $\{r\}=\{p^i\}$.
The matrix $U(x_\perp )$ is the $SU(N)$ matrix that parametrizes the
two dimensional ``pure gauge'' vector potential $\alpha^i(x_\perp)$
\begin{equation}
        \alpha^i(x_\perp )=
        {i\over g}U(x_\perp )\partial^i 
        U^\dagger(x_\perp )\nonumber
\end{equation}
The auxiliary functions $\gamma^i_+, v^i_\pm$ are all determined in
terms of one vector function. 
Choosing this independent function as $v^i_-$ we have
\begin{align}
        v^{i}_{+,r}&= \bigg[T^{ij} -L^{ij} \bigg]
        \bigg[t^{jk} -l^{jk}\bigg] v^{k}_{-,r} \\
        \gamma^{i}_{+,r}&= 2 D^{i} \bigg[{D^j\over D^2} -
        {\partial^j\over \partial^{2}}\bigg]
        \bigg[t^{jk} -l^{jk}\bigg]v^k_{-,r} 
\end{align}
where we have defined the projection operators
\begin{alignat}{2}
        T^{ij} & \equiv \delta^{ij} - 
        {D^i D^j \over D^2}, & 
        \qquad L^{ij} & \equiv 
        {D^i D^j \over D^2} \nonumber\\
        t^{ij} & \equiv  \delta^{ij} - 
        {\partial^i \partial^j \over \partial^2}, &
        \quad l^{ij} & \equiv  
        {\partial^i \partial^j \over \partial^2}
\label{eq:tproj}
\end{alignat}
The proper normalization of the eigenfunctions requires $v^i_-$ to be
chosen as complete set of eigenfunctions of the two dimensional
Hermitian operator $O^{-1}$ 
\begin{multline}
        \qquad [(t-l)O^{-1}(t-l)]_{ab}^{ij}(x_\perp,y_\perp)= \\
        <x_\perp|\delta_{ab}^{ij} - 
        2 \bigg[[\partial^i{1\over\partial^2}-D^i{1\over D^2}]
        S^{-1}[{1\over\partial^2}\partial^j-{1\over D^2}D^j]
        \bigg]_{ab}|y_\perp> \qquad
\end{multline}
such that
\begin{equation}
        \int d^2 r_\perp  v^{i}_{-,r,a}(x_\perp ) 
        v^{\ast j}_{-,r,b}(y_\perp ) = 
        {1\over 4\pi |p^-|}[O^{-1}]_{ab}^{ij} (x_\perp ,y_\perp )
\label{ort} 
\end{equation}

The rotational scalar operator $S$ is
\begin{equation}
        S = {1\over D^2}+2 [{\partial^i\over\partial^2}-
        {D^i\over D^2}][{\partial^i\over\partial^2}-
        {D^i\over D^2}]=
        {1\over D^2}-2{1\over \partial^2}
        \partial\alpha{1\over D^2} +
        2{1\over D^2}D\alpha{1\over\partial^2}
\end{equation}
  
For further use we also need the expression for the $a^-$ component of
the fluctuation field. Using the explicit expression for the operator $K$
from \cite{moap} we get from eq.(\ref{tildea})
\begin{multline}
        a^-(x^-,x_\perp,p^-) = 
        \tilde a^- -\theta(x^-) \int_{x^-}^\infty
        dy^-D^i(a^i-\gamma^i_+) \\
        -\theta(-x^-)\bigg[\int_0^\infty dy^-D^i(a^i-\gamma^i_+)
        +\int_{x^-}^0 dy^-\partial^ia^i\bigg] 
        + 2ip^-\eta[{D^i\over D^2} - {\partial^i\over \partial^2}]
        (t-l)^{ij}v^j_-(x_\perp)
\label{a-}
\end{multline}
We note that this expression differs by a $x^-$-independent constant
from the one given in
\cite{moap}. The reason is that in \cite{moap} a constant have been 
subtracted from
$a^-$ such that $\int_{-\infty}^{+\infty}dx^-a^-(x^-)=0$. This
corresponds to the symmetric definition of the integral in
eq.(\ref{a-}). This is incorrect, since it violates the residual gauge
fixing
$\partial^ia^i(x^-\rightarrow -\infty)$ at the one loop level. 
We will see this explicitly later in this section. At any rate,
straightforward albeit somewhat tedious calculation gives the result
eq.(\ref{a-}) and this is the expression that will be used in the rest
of this paper.
 
So far the formulae presented in this section 
(except for the corrected expression for $a^-$ eq.(\ref{a-}))
are identical to those that appear in \cite{moap}
with the only difference that the background charge 
density $\rho$ has been substituted by the background 
field via $\rho=-\partial^i\alpha^i$. 
Now however we will take a different route.
Our aim is to calculate the order $O(\alpha_s\ln 1/x)$ correction
to the background chromoelectric field eq.(\ref{chrom}) directly, rather than 
to the background charge density.
According to the discussion in the previous section (see also
\cite{moap}), we are therefore interested in the following two
quantities
\begin{align}
        \alpha_s\ln 1/x\: \chi_{ab}^{ij}(x_\perp,y_\perp) 
                &= <a_{a}^i(x_\perp,x^-\rightarrow\infty,
                        x^+)a_b^j(y_\perp,y^-\rightarrow\infty, x^+)> \\
        \alpha_s\ln 1/x\: \sigma_a^i(x_\perp)
                &= <a_a^i(x_\perp,x^-\rightarrow\infty,x^+)>
\end{align}
It should be noted that, since the background is static, none of the
quantities defined above depend on $x^+$.

\subsection{The real part - the mean fluctuation}

It is a straightforward matter to calculate $\chi^{ij}_{ab}$.
Recall that we need this quantity to order $g^2$. The fluctuation
fields $a^\mu$ are formally of order $g$ themselves, and therefore to
calculate the mean fluctuation we do not have to include loop
corrections.
Examining the expression for the general solution eq.(\ref{solut}) we
see that it contains oscillating pieces, which do not contribute
to the value of the field at infinity as well as the $\gamma_+$ piece,
which does not vanish at infinity and, therefore, determines the
distribution of the vector potential there.
\begin{equation}
        \chi_{ab}^{ij}(x_\perp,y_\perp)=4\pi\int d p^- 
                <\gamma^{i}_{+,a}(x_\perp,p^-)\gamma^{j}_{+,b}(y_\perp,-p^-)>
\end{equation}
Using the explicit expressions for $\gamma_+^i$ we find after some
trivial algebra
\begin{equation}
        \chi_{ab}^{ij}(x_\perp,y_\perp)=
        2<x_\perp|\{{D^i\over D^2}[D^2-S^{-1}]{D^j\over D^2}\}_{ab}|y_\perp>
\label{chif}
\end{equation}

We now want to compare this with the corresponding result of \cite{moap}.
The induced charge density $\delta\rho$ in \cite{moap} is 
\begin{equation}
        \delta\rho=\delta\rho_1+\delta\rho_2
\end{equation}
with
\begin{multline}
\label{rho11}
        \delta\rho_{1a}(x_\perp ) =  f_{abc} \alpha_{b}^{i}(x_\perp ) 
        \Bigg[a_{c}^{i}(x^-=0)-\int_0^{\infty}dx^- 
        \partial^+ a^i_c(x^-)\Bigg]  \\
        - {{1}\over{2}} f_{abc} \partial^i  \alpha^{i}_{b}(x_\perp ) \int dy^+ 
        \Bigg[\theta (y^+ - x^+) - \theta (x^+ - y^+) \Bigg] 
        a^{-}_{c}(y^+,x_\perp ,x^-=0)
\end{multline}
and
\begin{multline}
\label{rho21}
        \delta \rho_{2a}(x_\perp ) = 
        f_{abc} \int d x^- [\partial^+ a_{b}^{i}(x) ]a_{c}^{i}(x) \\ 
        - {1\over{2}} \partial^i  \alpha^{i}_{b}(x_\perp ) 
        \int\! dy^+ a^{-}_{c}(y^+,x_\perp ,x^-=0)
        \int\! dz^+ a^{-}_{d}(z^+,x_\perp ,x^-=0) \\ 
        \times \Bigg[f_{ace}f_{bde}\theta (z^+ -x^+)\theta (x^+ -y^+) + 
        f_{abe} f_{cde}\theta (x^+ -z^+) \theta (z^+ -y^+) \Bigg]
\end{multline}
Only $\delta\rho_1$ contributes to $\chi$.
Substituting the expressions for $a^i$ and $a^-$ into eq.(\ref{rho11})
we find
\begin{equation}
        \delta\rho_1=
        -2(\partial D)
        [{D\over D^2}-{\partial\over\partial^2}]
        (t-l)v_-
\end{equation}
Thus, we obviously have
\begin{equation}
        \gamma_+^i=-D^i{1\over \partial D}\delta\rho_1
\label{relat}
\end{equation}
This reproduces exactly eq.(\ref{corresp1}). 
Obviously the relation between $\chi_{ab}^{  }$ and $\chi_{ab}^{ij}$,
eq.(\ref{indcor}) is also reproduced by this result. 

We note that our result for $\chi_{ab}^{  }$ is somewhat different than the
one presented in \cite{moap}. As discussed before this is due to an
incorrect treatment of the $x^-$-independent component of $a^-$ in
\cite{moap}.

\subsection{The virtual part - the average value of the field}

We now proceed to calculate the virtual part of the evolution kernel. 
For this purpose we have to calculate the zero frequency part of the
$\{ij\}$ and $\{i-\}$ components of the fluctuation propagator.
The calculation of the $\{ij\}$ at zero frequency is straightforward.
The result is
\begin{multline}
        \lim_{p^-\rightarrow 0} 
        G_{ab}^{ij}(x^-,y^-;x_\perp,y_\perp,p^-)
        \equiv \lim_{p^-\rightarrow 0} <a_a^i(x^-,x_\perp,p^-)
        a_b^j(y^-,y_\perp,p^-)>=  \\
        -i\delta^{ij} \delta(x^--y^-)
        \Big [\theta(x^-)
        <x_\perp|({1\over D^2})_{ab}|y_\perp>+
        \theta(-x^-)<x_\perp|({1\over \partial^2})_{ab}|y_\perp>
        \Big]
\end{multline}
The $\{i-\}$ component is then calculated immediately using this
result, eq.(\ref{a-}) and the fact noted earlier that the field
$\tilde a$ decouples from $a^i$.
The result is
\begin{multline}
        \lim_{p^-\rightarrow 0}
        G_{ab}^{i-}(x^-,y^-;x_\perp,y_\perp,p^-)
        \equiv \lim_{p^-\rightarrow 0} <a_a^i(x^-,x_\perp,p^-)
        a^-_b(y^-,y_\perp,p^-)>= \\
        i\theta(x^--y^-)
        \Big[\theta(x^-)
        <x_\perp|({D^i\over D^2})_{ab}|y_\perp>+
        \theta(-x^-)<x_\perp|({\partial^i\over
        \partial^2})_{ab}|y_\perp>
        \Big] 
\end{multline}

\begin{fmffile}{moap1pics}
\unitlength=1mm
\gdef\T#1#2#3#4{
\begin{fmfgraph*}(40,16)
\fmfpen{.6thin}
\fmfi{wiggly}{(0,.5h) -- (.6w,.5h)}
\fmfi{wiggly}{fullcircle scaled .4w shifted (.8w,.5h)}
\def\V##1##2##3{
        \fmfiv{dec.shape=circle, dec.size=3, lab.angle=##3, 
        label=\noexpand\texttt{\noexpand\small ##1}}{##2}}
\V{#1}{(0,.5h)}{95}
\V{#2}{(.6w,.5h)}{140}
\V{#3}{(.6w,.5h)}{40}
\V{#4}{(.6w,.5h)}{-40}
\end{fmfgraph*}}
\begin{figure}[t]
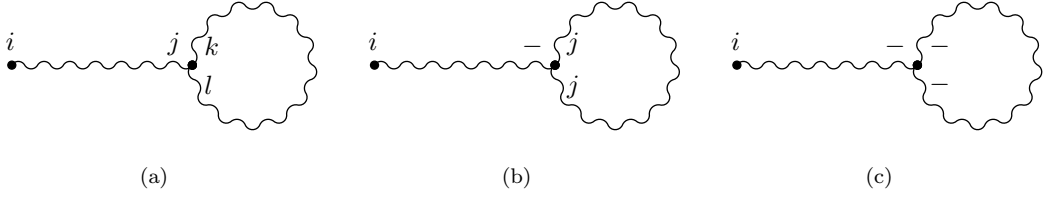

\centering
\mbox{  \subfigure[]{\T{$i$}{$j$}{$k$}{$l$}}\qquad
        \subfigure[]{\T{$i$}{$-$}{$j$}{$j$}}\qquad
        \subfigure[]{\T{$i$}{$-$}{$-$}{$-$}}}
\caption{\small One loop tadpole diagrams contributing to
$\sigma^i_a$. The tadpole is calculated at $x^-\rightarrow\infty$.}
\end{figure}
\end{fmffile}

We are now ready to calculate $\sigma_a^i$. 
It is given by the one loop tadpole diagrams of Fig. 1.
The vertex 1c comes from the expansion of the Wilson line
term in the action to third order in the fluctuation. 
The separate contributions of the diagrams can be written in terms of
the fluctuation propagator $G^{\mu\nu}\equiv<a^\mu a^\nu>$ in  the
following form
\begin{align}
\text{1a} &=    \frac{i}{2}\int d y^-d^2 y_\perp 
                G^{ij}_{ab}(x^-, y^-, x_\perp,y_\perp, p^-=0) 
                \epsilon^{jk}D^k_{bc}f_{cde} \epsilon^{mn}
                G^{mn}_{de}(y^-,y^-;y_\perp,y_\perp,y^+,y^+) 
                \nonumber \\
\text{1b} &=            - i \int d y^-d^2 y_\perp 
                G^{i-}_{ab}(x^-, y^-, x_\perp,y_\perp, p^-=0)
                f_{bcd}{\partial}^{+}_{\tilde{y}^- = y^-}
                G^{jj}_{cd}(y^-,\tilde{y}^-;y_\perp,y_\perp,y^+,y^+) 
        \nonumber \\
\begin{split}
\text{1c} &=    \frac{i}{N_c} \int d y^-d^2 y_\perp d y^+ d w^+ d z^+ \\
          &     \qquad\qquad\times 
                \delta (y^-) ({\partial}^{i}{\alpha}^{i}_{b}(y_\perp)) 
                G^{i-}_{ac}(x^+, y^+,x^-, y^-, x_\perp,y_\perp) 
                G^{--}_{de}(w^+, z^+,y^-, y^-, y_\perp,y_\perp)
           \\
          &     \qquad\qquad\quad\times
                \bigg[ 
                \theta (z^+ - y^+)\theta (y^+ - w^+) f_{bef}f_{cdf} -
                \theta (y^+ - z^+)\theta (z^+ - w^+) f_{bcf}f_{def}
                \bigg]
\end{split}
\end{align}

The diagram Fig.~1a corresponds directly to the second term in
eq.(\ref{indfield}).
For this diagram we immediately find
\begin{equation}
        \delta\sigma^i_{a(1)}(x)=
        -{1\over 2}\epsilon^{ij}
        \left[{D^j\over D^2}\right]_{ab}\negthickspace (x,y)
        f_{bcd}\epsilon^{kl}\chi^{kl}_{cd}(y,y)
\label{finals1}
\end{equation}

The diagramms Fig.~1b and 1c correspond to
the first term in eq.(\ref{indfield}) and can be written as
\begin{equation}
        \delta\sigma^i_{a(2)}(x)=
        -{D^i\over D^2}<\delta\rho_2>
\label{stum}
\end{equation}
with $\delta\rho_2$ (cf eq.(\ref{rho21})):
\begin{multline}
        <\delta\rho_2>_a = 
        f_{abc} \int d x^- 
        < (\partial^+ a_{b}^{i}(x) )a_{c}^{i}(x)> 
        + \, {1\over{2}} \, 
        (f_{ace}f_{bde} - \frac{1}{4} f_{abe}f_{cde}) \,
        \partial^i \alpha^i_b (x_\perp ) \\
        \times
        \int\! \frac{d\lambda}{\lambda +i\epsilon} 
        dp^- \frac{1}{(p^-)^2} 
        <a^{-}_{c}(p^-, x_\perp ,x^-=0) 
        a^{-}_{d}(-p^-, x_\perp ,x^-=0)>
\label{finals}
\end{multline}

Using the results for the equal time propagators from \cite{moap} we
obtain
\begin{equation}
\begin{split}
        <\delta\rho_2>_a 
        &= 
        - {1\over 2}\, (f_{ace}f_{bde} - \frac{1}{4} f_{abe}f_{cde})\, 
        \partial^i \alpha^i_b (x_\perp) \\
        &\qquad\qquad\times <x_\perp| {1\over \partial^2}+
        {1\over 2}\mu{1\over D^2}\mu 
        -2\left [{1\over \partial^2}\alpha D+{\mu\over 2} \right ] 
        {1\over D^2}S^{-1}{1\over D^2}
        \left [D\alpha{1\over \partial^2}
        +{\mu\over 2}\right ]|x_\perp>_{cd} \\
        &\quad+f_{abc}<x_\perp|
        \left [t^{ij}-l^{ij}
        -2\alpha^i\partial^j{1\over \partial^2} \right]
        \left [\delta^{jk}-2({\partial^j\over\partial^2}
        -{D^j\over D^2})S^{-1} ({\partial^k\over\partial^2}
        -{D^k\over D^2})\right ] \\
        &\hspace{74mm}
        \times \left [T^{ki}-L^{ki}
        -2{1\over \partial^2}\partial^k\alpha^i\right ]|x_\perp>_{bc}\\
        &\quad +R^a(x_\perp)
\end{split}
\label{well}
\end{equation}
with
\begin{equation}
\begin{split}
        R^a(x_\perp)
        &=      
        f_{abc}\int d^2y_\perp d^2z_\perp 
        {d^2p_\perp d^2k_\perp\over (2\pi)^4}
        {p_\perp^2\over p_\perp^2-k_\perp^2}
        e^{ip_\perp (x_\perp-y_\perp)-ik_\perp (x_\perp-z_\perp)}\\
        &\qquad\times \Big \{
        <y_\perp|\left [t^{ij}-l^{ij}\right]
        \left [\delta^{jk}
        -2({\partial^j\over\partial^2}-{D^j\over D^2})S^{-1}
        ({\partial^k\over\partial^2}-{D^k\over D^2})\right ]
        \left [t^{ki}-l^{ki}\right ]|z_\perp> \\
        &\hspace{20mm}
        -U(x_\perp)<y_\perp|U^\dagger\left [T^{ij}-L^{ij}\right]
        \left [\delta^{jk}
        -2({\partial^j\over\partial^2}-{D^j\over D^2})S^{-1}
        ({\partial^k\over\partial^2}-{D^k\over D^2})\right ] \\
        &\hspace{85mm}\times
        \left [T^{ki}-L^{ki}\right ]U|z_\perp>
        U^\dagger(x_\perp) \Big \}_{bc}
\end{split}
\label{R}
\end{equation}
Here the singularity in the integrand at $p_\perp^2=k_\perp^2$ has to be
understood in the sense of the principal value 
$${1\over p_\perp^2-k_\perp^2}={p_\perp^2-k_\perp^2
\over(p_\perp^2-k_\perp^2)^2+\epsilon^2}.$$

Our final result for the induced field is
given by the sum of eq.(\ref{finals1}) and eq.(\ref{stum})
(supplemented by eqs.(\ref{well},\ref{R})).

\section{Conclusions}

To summarize, the final results of this paper are eqs.(\ref{chif}) and
(\ref{stum},\ref{well}).
They supercede the corresponding results of \cite{moap} and
\cite{soap}.

We now want to comment on this result. The first thing to observe is
that  the dangerous
denominator $\partial D$ does not appear in these expressions. 
The Gribov problem mentioned earlier
therefore does not affect our calculation, at least
to order $\alpha_s$.

The result for the induced field differs from the corresponding formulae
in \cite{moap} and \cite{soap} in two ways. 
One
reason is the improved
treatment of $a^-$ relative to \cite{moap}. 
Now we are in the position to understand why the expression for $a^-$
used in \cite{moap} is inconsistent with the residual gauge fixing.
In the previous section we have calculated the induced vector
potential 
far at infinity
$x^-\rightarrow\infty$. It is not much more difficult to calculate
it everywhere in space. Diagrammatically it is given by
the same diagrams as Fig.1 except the coordinate on the free end of
the propagator is some finite $x^-$. The difference in the analytic
expressions eq.(\ref{stum}) is that the surface charge density
$\delta\rho_2$ is substituted by the local charge density integrated
up to the longitudinal coordinte $x^-$
\begin{equation}
-\theta(x^-){D^i\over D^2}\int_{-\infty}^{x^-}dy^-<\delta j^+_2(y^-)>+
-\theta(-x^-)
{\partial^i\over \partial^2}\int_{-\infty}^{x^-}dy^-<\delta j^+_2(y^-)>
\label{stump}
\end{equation}
This expression makes it explicit that the induced field vanishes at
$x^-\rightarrow-\infty$.
Therefore, our calculation clearly preserves the residual gauge
condition $\partial^ia^i(x^-\rightarrow-\infty)=0$. 
However if we were to subtract the zero momentum piece from the field
$a^-$ as done in \cite{moap}, the integration limits in eq.(\ref{a-})
would become symmetric $\int_x^\infty \rightarrow {1\over
2}(\int_x^\infty+\int_x^{-\infty})$. 
The effect of this would be that $G^{i-}(x^-,y^-)$ would not vanish at
$x^-\rightarrow -\infty$.
It is then obvious that we would have $\partial^ia^i(x^-\rightarrow
-\infty)\ne 0$. 
The expression obtained in the present paper does not suffer from this
problem. 
It is consistent with the perturbative $i\epsilon$ prescription for
regulating the $1/p^+$ gauge pole used in the earlier work
\cite{kovchegov}.

Another difference between our present result and \cite{moap} is the
appearance of $D^2$ rather than $\partial D$ and $D\partial$ in the
denominators in eqs.(\ref{finals1},\ref{stum}).
This deserves an explanation. 
This is also related to another point we want to address.
Comparing eq.(\ref{stum}) with eq.(\ref{indfield}) one could wonder
whether the present method of calculation of $\chi^{ij}$ is consistent
with the two step procedure of \cite{moap, soap}. 
It may look like the relation between the induced field and the
induced charge density we obtained here (eq.(\ref{stum})) is different
from the equation eq.(\ref{indfield}) which was used in the previous
work.
This however is not the case.
The reason is that the $O(g^2)$ induced charge density
$\delta\tilde\rho_2$ which appears in eq.(\ref{indfield}) is not quite
the same as $<\delta\rho_2>$ in eq.(\ref{stum}). 
The $\delta\tilde\rho_2$ was defined as complete $O(g^2)$ contribution
to the average of induced density.
In other words 

\begin{equation}
\delta\tilde\rho_2=<\delta\rho_1+\delta\rho_2>
\label{tilnottil}
\end{equation}
with $\delta\rho_{1,2}$ defined in eqs.(\ref{rho11},\ref{rho21}).
As we discussed above, the fluctuating part of the operator
$\delta\rho_1$ is of $O(g)$ and therefore indeed $\delta\tilde\rho_1$
can be identified with $\delta\rho_1$. 
However, the vacuum average of $\delta\rho_1$ is $O(g^2)$ and does
contribute in eq.(\ref{tilnottil}).
It can be shown that 
\begin{equation}
<\delta\rho_{1a}>=f_{abc}\alpha_b^i<a^i_c(x^+\rightarrow\infty)>
\label{tr}
\end{equation}
This extra contribution turns $\partial^i\delta\alpha_2^i$ into 
$D^i\delta\alpha_2^i$ in the
second equation in eq.(\ref{solp}) if we use $\delta\rho_2$ rather
than $\delta\tilde\rho_2$ in its right hand side.
Taking account of this we see that the procedure described in section
2 is consistent with
eqs.(\ref{chif},\ref{stum},\ref{well}).

In \cite{moap} it was assumed that $<\delta\rho_1>=0$ and thus the
extra contribution of eq.(\ref{tr}) was overlooked.
This lead to an apparent noncancellation  of spurious factors ${1\over
\partial D}$ which as we see now, do indeed cancel in the final
result.

Importantly, the corrections we find vanish in the limit of weak field
considered in \cite{jklw} and also in the double logarithmic limit,
where the field is considered not necessarily weak but slowly varying
in the transverse plane \cite{soap}.
This can be seen in the following way. 
Comparing eq.(\ref{a-}) to the appropriate expression in \cite{moap}
we find that the difference between the two is proportional to
$\rho$. 
In the weak field limit one only needs to know $a^-$ to order $1$ and
therefore the correction is unimportant. 
For slowly varying fields all terms proportional to $\rho$ are also
negligible. 
Therefore, the real part - $\chi^{ij}$ - in these two limits is
insensitive to the correction we found here. 
The virtual part - $\sigma^i$ - does not contribute at all in the DLA
limit. 
In the weak field limit the correction is negligible since
$<\delta\rho_1>=O((\alpha^{i})^2)$ and one only needs $\delta\rho$ to
order $\alpha^i$.

\pagebreak
{\bf Acknowledgements}
We are greatful to J. Jalilian-Marian, L. McLerran and H. Weigert
for useful discussions.
The work of J.G.M. is supported by PRAXIS
XXI/BD/11277/97 grant (Subprograma
Ci\^encia e Tecnologia do 2$^{\underline o}$ Quadro
Comunit\'ario de Apoio --- Portugal).
The work of A.K. is supported by PPARC
Advanced fellowship.


\begin{thebibliography}{99}


\bibitem{dglap} see e.g. Yu.~Dokshitzer, V.~Khoze, A.~Mueller and 
S.~Troyan, "Basics of Perturbative QCD", ed. Frontiers, 1991.

\bibitem{bfkl}E.A.~Kuraev, L.N.~Lipatov and V.S.~Fadin, {\it{Sov.\
Phys.\ JETP}}\ {\bf{45}} (1977) 199; Ya.Ya.~Balitsky and
L.N.~Lipatov, {\it{Sov.\ J.\ Nucl.\ Phys.}}\ {\bf{28}} (1978) 22.

\bibitem{levin} A.L.~Ayala, M.B.~Gay Ducati and E.M.~Levin,
hep-ph/9706448, {\it  Nucl. Phys.} {\bf B511} (1998) 355; E.~Levin,
Talk at "Continuous Advances in QCD", Minneapolis, April 1998,
hep-ph/9806434; see also E.~Levin, hep-ph/9709226 and E.~Gotsman,
E.~Levin and U.~Maor, hep-ph/9712517, {\it Phys. Lett.} {\bf
B425} (1998) 369-374.

\bibitem{muller} A.~Mueller, hep-ph/9710531, {\it Eur. Phys. J.} {\bf
A1} (1998) 19.

\bibitem{levin2} E.~Gotsman, E.M.~Levin and U.~Maor, hep-ph/9606280, 
{\it Nucl. Phys.} {\bf B493} (1997) 354.

\bibitem{glr}
L.V.~Gribov, E.M.~Levin and M.G.~Ryskin, {\it{Phys.\ Rep.}}\ {\bf{100}}
(1981).

\bibitem{mq}  A.H.~Mueller and J.W.~Qiu, {\it Nucl. Phys.} {\bf B268}
(1986) 427.

\bibitem{mueller} A.H.~Mueller, {\it Nucl. Phys.} {\bf B335} (1990)
115.

\bibitem{mv} L.~McLerran and R.~Venugopalan, \mbox{hep-ph/9309289}, 
{\it{Phys. Rev.}} {\bf D49} (1994) 2233; \mbox{hep-ph/9311205}, 
{\bf D49} (1994) 3352.

\bibitem{jkmw} J.~Jalilian-Marian, A.~Kovner, L.~McLerran and
H.~Weigert,  hep-ph/9606337, {\it Phys. Rev.} {\bf D55} (1997) 5414.

\bibitem{jklw} J.~Jalilian-Marian, A.~Kovner, A.~Leonidov and
H.~Weigert, hep-ph/9701284, {\it Nucl.~Phys.} {\bf B504} (1997) 415;
J.~Jalilian-Marian, A.~Kovner, A.~Leonidov and H.~Weigert,
\mbox{hep-ph/9706377}, {\it Phys. Rev.} {\bf D59} (1999) 014014.
 
\bibitem{moap} J.~Jalilian-Marian, A.~Kovner and H.~Weigert,
hep-ph/9709432, {\it Phys. Rev.} {\bf D59} (1999) 014015.

\bibitem{soap} J.~Jalilian-Marian, A.~Kovner, A.~Leonidov and
H.~Weigert, hep-ph/9807462, {\it Phys. Rev.} {\bf D59} (1999) 034007, 
Erratum-ibid.{\bf D59} (1999) 099903.

\bibitem{jw} J.~Jalilian-Marian, X-N~Wang, hep-ph/9902411. 

\bibitem{gv} R.~Gavai and R.~Venugopalan, hep-ph/9605327, 
{\it Phys. Rev.} {\bf D54} (1996) 5795-5803. 


\bibitem{kv} A.~Krasnitz and R.~Venugopalan, hep-ph/9809433.

\bibitem{kovchegov} Yu.~Kovchegov, hep-ph/9701229, {\it
Phys. Rev.} {\bf D55} (1997) 5445-5455. 


\end{thebibliography}
\end{document}